\documentstyle[prl,aps,multicol,psfig]{revtex}
\tighten
\begin{document}
\draft
\preprint{\today}

\title{Level-spacing distribution of a fractal matrix}
\author{D.E. Katsanos$^{1}$ and S. N. Evangelou$^{1,2}$}
\address{$^1$Department of Physics, University of Ioannina, Ioannina 45 110,
Greece \\
$^2$Department of Physics, University of Lancaster, 
Lancaster LA1 4YB, UK \\}
\date{\today}
\maketitle

\begin{abstract}
We diagonalize numerically a Fibonacci matrix with fractal 
Hilbert space structure of dimension $d_{f}=1.8316...$ We show that the
density of states is logarithmically normal while the corresponding 
level-statistics can be described as critical since the nearest-neighbor 
distribution function approaches the intermediate semi-Poisson curve. 
We find that the eigenvector amplitudes of this matrix are 
also critical lying between extended and localized.
\end{abstract}


The numerical diagonalization of one-electron Hamiltonians 
in discrete space tight-binding lattices has been proved a very useful 
tool to treat the effect of disorder on the motion of quantum particles,
alternatively  to field theoretic methods. 
In particular, the direct computation of eigenvalues and eigenvectors 
combined with finite-size scaling techniques can answer questions  
about Anderson localization due to disorder \cite{1}, 
quantum chaos in the energy levels or wave functions \cite{2,3}, etc. 
The difficulties become immense when 
one wants to treat many-body problems in the same framework 
since the dimension of the Hilbert space increases dramatically 
(exponentially) with the number of electrons. However, also in this case 
the nearest-neighbor hopping bounded kinetic energy 
and the two-body character of the interaction guarantee 
that the Hamiltonian matrix structure is very sparse, 
often described as multifractal in the adopted Hilbert space.  
For example, in ref. \cite{4} multifractal exponents $D_{q}$ 
were computed to characterize the  Hilbert space  
structure of interacting many-body Hamiltonians.

We shall treat a much simpler, albeit interesting problem,  
with a given Hamiltonian which consists of zeroes and ones,
neither periodic nor random but with a simple fractal structure.
We construct and diagonalize a Fibonacci matrix of order 
$n$ of size $N_{n}\times N_{n}$, with $N_{n}=N_{n-1}+N_{n-2}$, 
$N_{0}=N_{-1}=1$.  
This matrix represents a self-similar fractal object itself and was
introduced in ref. \cite{5} in order to obtain the ground state of the 
square Ising antiferromagnet in a maximum critical field.
The Fibonacci matrix is a transfer matrix 
connecting the possible ground states of the $n\times m$ 
and $n\times (m+1)$ square lattices. For size $N_{n}$ it has 
the block form
\begin{equation}
\label{matrix}
F_{n}=\left (\begin{array}{cc} F_{n-1} & G_{n-1} \\
	       G_{n-1}^{T} & 0
	\end {array}\right)
\end{equation}
where $G_{n-1}$ represents the $N_{n-1}\times N_{n-2}$ submatrix 
of $F_{n-1}$ (its upper left corner), $G_{n-1}^{T}$ its transpose
and $0$ denotes the zero $N_{n-2}\times N_{n-2}$ matrix. 
The sequence of fractal matrices for $n=1,2,...$ 
begins with the matrix of size $N_{1}=2$ 
\begin{equation}
\label{matrix1}
F_{1}=\left (\begin{array}{cc} 1 & 1 \\  
	       1 & 0
	\end {array}\right),
\end{equation}
size $N_{2}=3$  
\begin{equation}
\label{matrix2}
F_{2}=\left (\begin{array}{ccc} 1 & 1 & 1 \\  
	       1 & 0 & 1 \\
	       1 & 1 & 0
	\end {array}\right), 
\end{equation}
size $N_{3}=5$  
\begin{equation}
\label{matrix3}
F_{3}=\left (\begin{array}{ccccc} 1 & 1 & 1 & 1 & 1 \\  
	       1 & 0 & 1 & 1 & 0\\
	       1 & 1 & 0 & 1 & 1\\
	       1 & 1 & 1 & 0 & 0\\
	       1 & 0 & 1 & 0 & 0
	\end {array}\right), 
\end{equation}
$\it{etc}$ and the corresponding matrices for higher $n$'s are easily 
obtained from eq. (\ref{matrix}) successively. For example, the 
matrix $F_{10}$ of size $N_{10}=144$ is shown in fig. 1.

%%
%%-------------------fig 1
%%
\par
\vspace{.3in}
\centerline{\psfig{figure=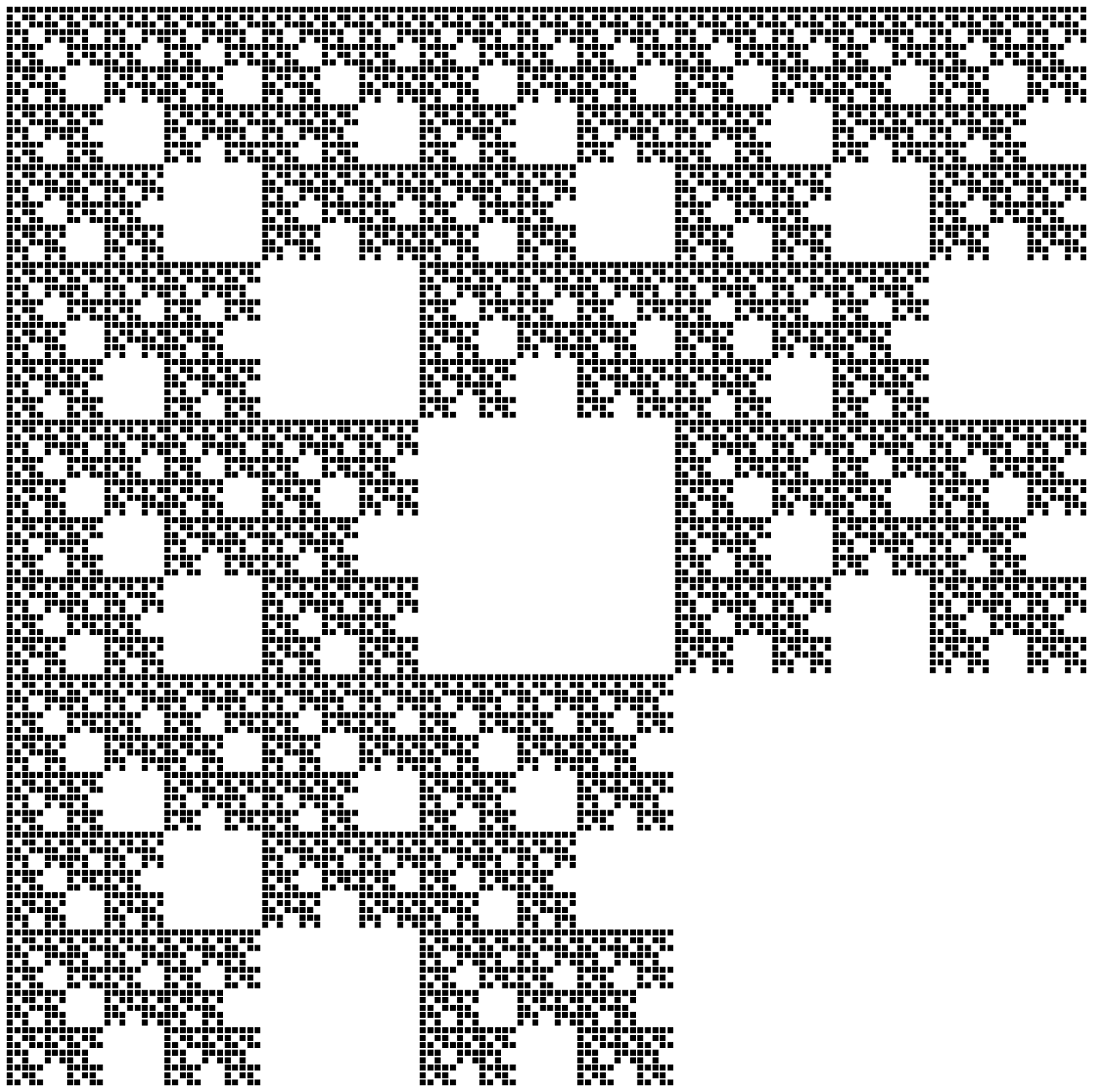,width=10.0cm}}
{\footnotesize{{\bf FIG. 1.} The fractal Fibonacci matrix $F_{10}$ of 
size $N_{10}=144$. The ones are denoted with black and the zeroes 
with white.}}
\vspace{0.3in}

In the Fibonacci matrix structure of fig. 1 we observe
no coverage of the full two-dimensional 
Hilbert space, with blocks within blocks,
so that the matrix is a simple fractal object.  
The fractal dimension of the Fibonacci matrix 
can be obtained form eq. (1) if we associate the mass $M_{n}$ 
measured by the number of $1$'s to the $F_{n}$ matrix of size $N_{n}$,
via the scaling relation
\begin{equation}
\label{frelation}
M_{n}=N_{n}^{d_{f}}.
\end{equation}
The corresponding asymptotic ($n\to \infty$) size
ratio $N_{n+1}/N_{n}=(1+\sqrt{5}))/2$ and mass ratio 
$M_{n+1}/M_{n}=1+\sqrt{2}$ for two successive generations $n$ and $n+1$ 
can enter eq. (5). The size ratio is obtained from two successive Fibonacci 
approximants (equals the ratio of the sides of two successive rectangles) 
and by definition corresponds to the golden mean $(1+\sqrt{5})/2$.
The mass ratio $M_{n+1}/M_{n}=1+\sqrt{2}$ is obtained from the 
of the series $M_{n}=M_{n-1}+2(M_{n-2}+M_{n-3}+M_{n-4}+...+M_{1}+2)$,
$n=2,3,..$, $M_{1}=3$ and the formation of 
$M_{n+1}/M_{n}=1+2(
{\frac {1} {M_{n}/M_{n-1}}}+
{\frac {1} {M_{n}/M_{n-2}}}+
...+
{\frac {1} {M_{n}/M_{1}}}+
{\frac {1} {M_{n}/2}})$ which in the limit of $n\to \infty$ where
${\frac {M_{n+1}} {M_{n}}}=
{\frac {M_{n}} {M_{n-1}}}\to x$,
${\frac {M_{n}} {M_{n-2}}}=
{\frac {M_{n}} {M_{n-1}}} {\frac {M_{n-1}} {M_{n-2}}}\to x^{2}$, etc,
becomes $ x=1+2({\frac {1} {x}}+
{\frac {1} {x^2}}+
...)$ 
and  by summing the geometric series we obtain $x=1+\sqrt{2}$. 
Therefore, the matrix of fig. 1 is a simple 
fractal (monofractal) object with dimension between $1$ and $2$  given by
\begin{equation}
\label{dimension}
d_{f}={\frac {log(1+\sqrt{2})} {log((1+\sqrt{5})/2)}}=1.8316...
\end{equation}
In order to obtain the residual entropy 
of the antiferromagnetic Ising model in maximum critical field
the leading eigenvalue of such Fibonacci matrices 
up to $F_{9}$ was estimated  in ref. \cite{5}.

Our purpose is entirely different. 
We shall consider the full spectrum of the Fibonacci matrix 
(eigenvalues and eigenvectors) and address questions related 
to Anderson localization and quantum chaos.  
We emphasise that we do not have in mind any
particular physical system described by this matrix
although such fractal matrices might be in general relevant for critical 
(between extended and localized) one-electron states
in disordered or quasiperiodic potentials or possibly
many-electrons states for interacting electrons in the presence 
of disorder \cite {4}.
We shall address the following questions
concerning the Fibonacci matrix: 

(1) What is the density of states? 

(2) What is the required `unfolding' procedure and the 
nearest-level spacing distribution function?

(3) What is the behavior of the corresponding eigenvectors?

%%
%%-------------------fig 2a, 2b
%%
\par
\vspace{0.3in}
\centerline{\psfig{figure=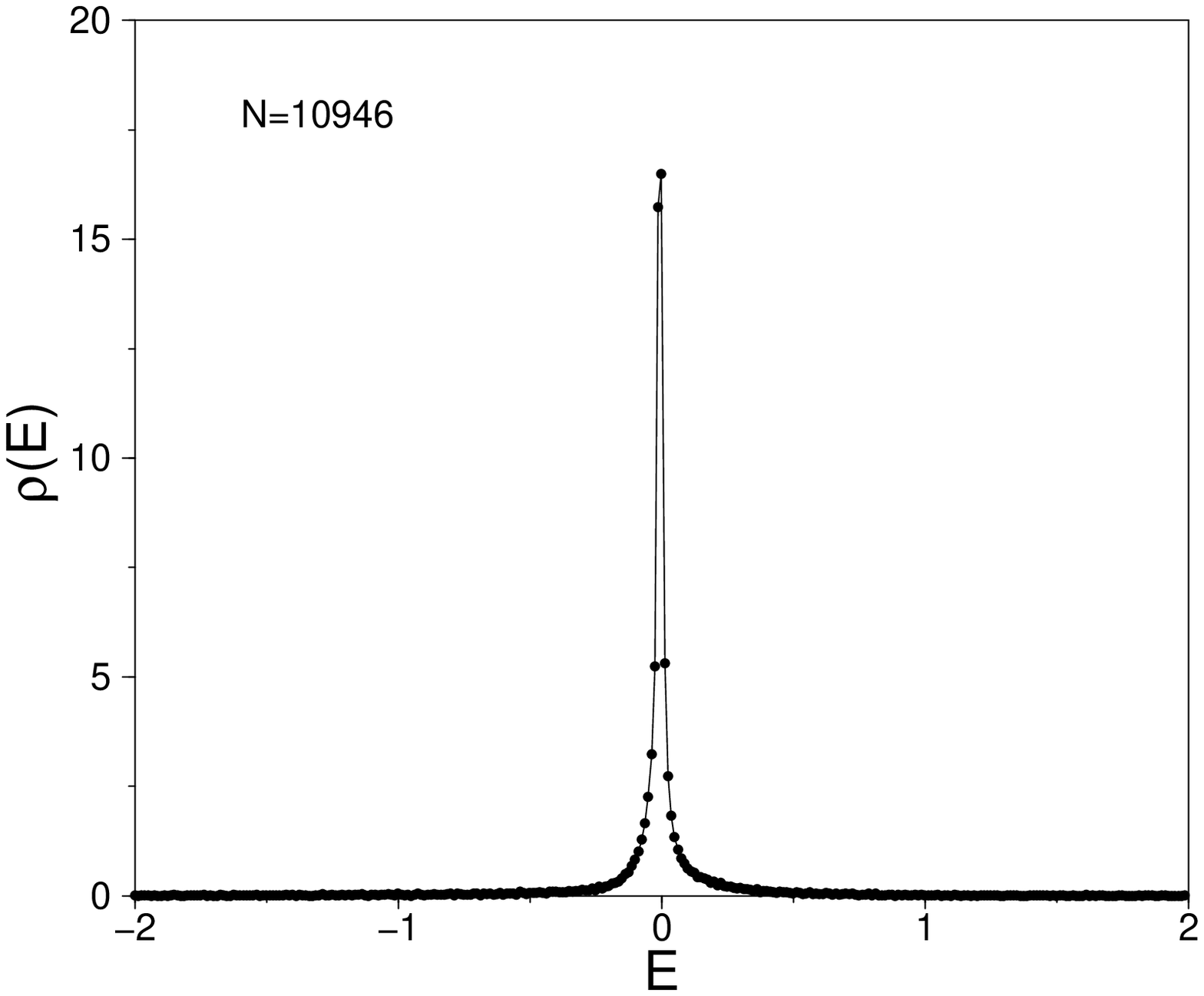,width=8.0cm}}
\vspace{.3in}
\centerline{\psfig{figure=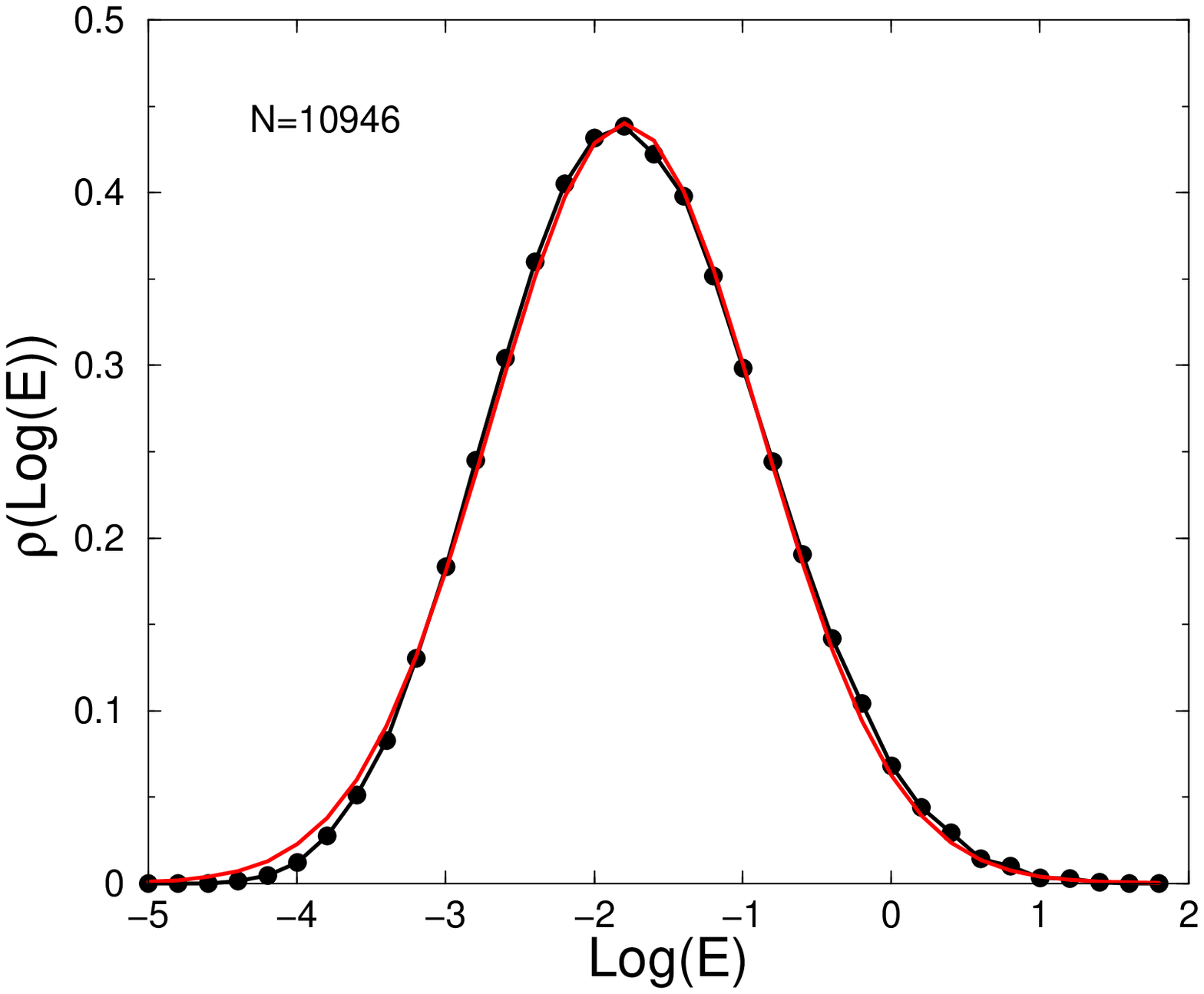,width=8.0cm}}
{\footnotesize{{\bf FIG. 2.} {\bf (a)} The density of states 
(full circles joined by a solid line) for the Fibonacci matrix $F_{19}$ 
of size $ N=N_{19}=10946$.
{\bf (b).} The corresponding density of states 
(full circles joined by a solid line) as a function of $\log(E)$, 
where $E$ is the absolute value of the energy. The 
simple Gaussian fit with mean $-1.79$ and standard deviation $0.91$
is also shown by the continuous line.}} 
\vspace{0.3in}

Firstly, we compute the density of states $\rho(E)$ as a  function
of the energy $E$ which is shown in fig. 2(a). It is seen to be 
dominated by a strong peak at the  origin $E=0$ and also has
very long tails which increase with the size. 
Instead, the distribution of $\rho(logE)$ shown in fig. 2(b),
where $E$ is the absolute value of the energy, leads to a Gaussian-like
function. From the Gaussian fit of fig. 2(b) we can conclude that 
the density of states $\rho(E)$ of the Fibonacci matrix 
follows a log-normal distribution. We have obtained similar
densities of states for all other sizes examined. 
Having answered the first question,
we proceed to compute the energy level-statistics 
in this system. In order to do so it is required to solve the problem 
of `unfolding' for the unusual density of states of fig.2 which demonstrates 
a  strong peak at the origin. 
It is impossible to obtain a meaningful level-statistics
without solving the `unfolding' problem which is required to make the 
density of states (particularly this unusually singular
density) constant, with mean-level spacing $\Delta=1$.
Our approach consists in `unfolding' the energy levels 
by using the density of fig. 2(b) instead of using that of fig. 2(a). 
In fact, this is essentially  a double unfolding procedure 
of the original spectrum, one going from fig. 2(a) to 2(b) and the second  
done for the data of fig. 2(b), so finally the mean level-spacing 
becomes $\Delta=1$.

%%
%%-------------------fig 3a,b
%%
\par
\vspace{.3in}
\centerline{\psfig{figure=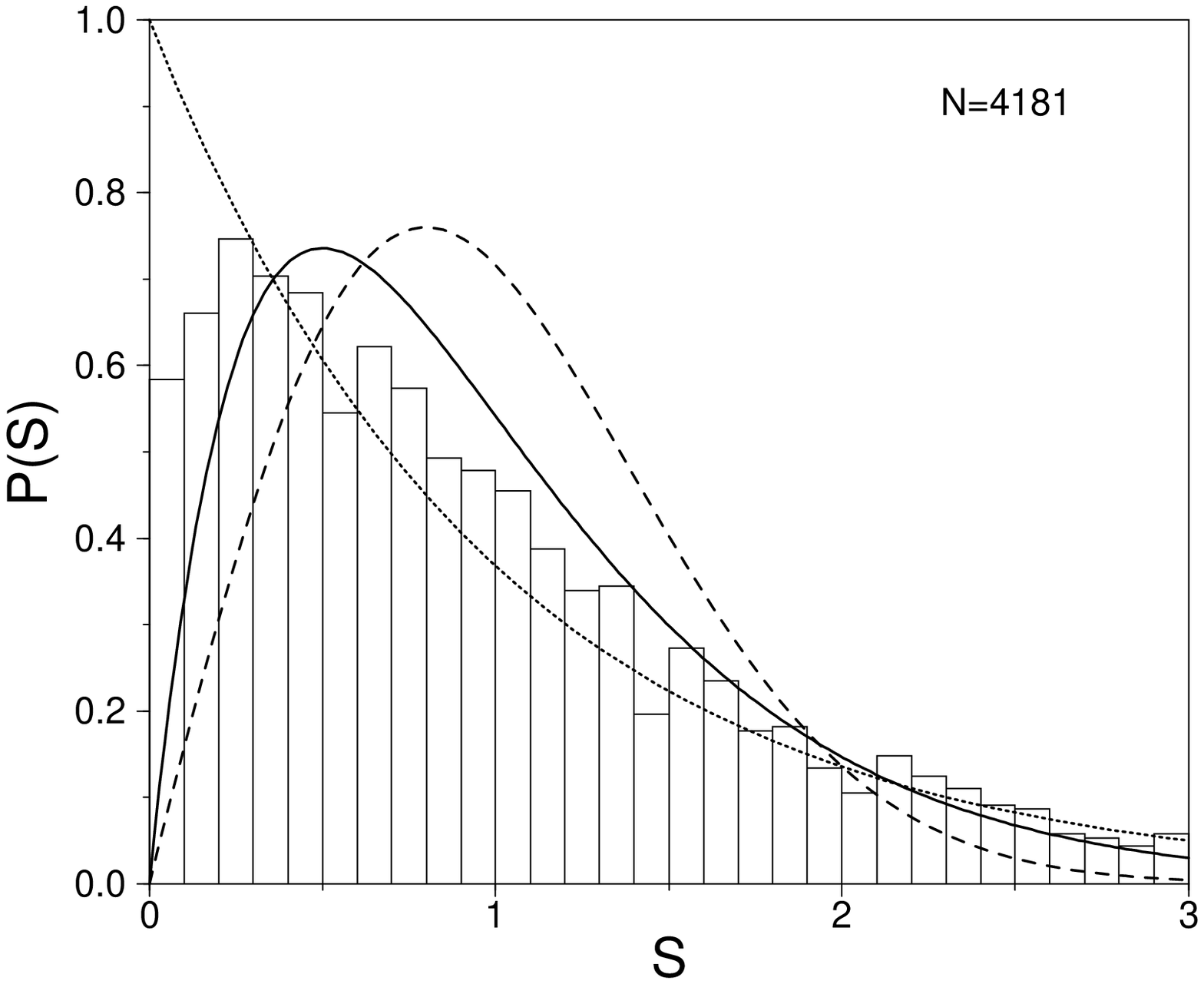,width=8.0cm}}
\centerline{\psfig{figure=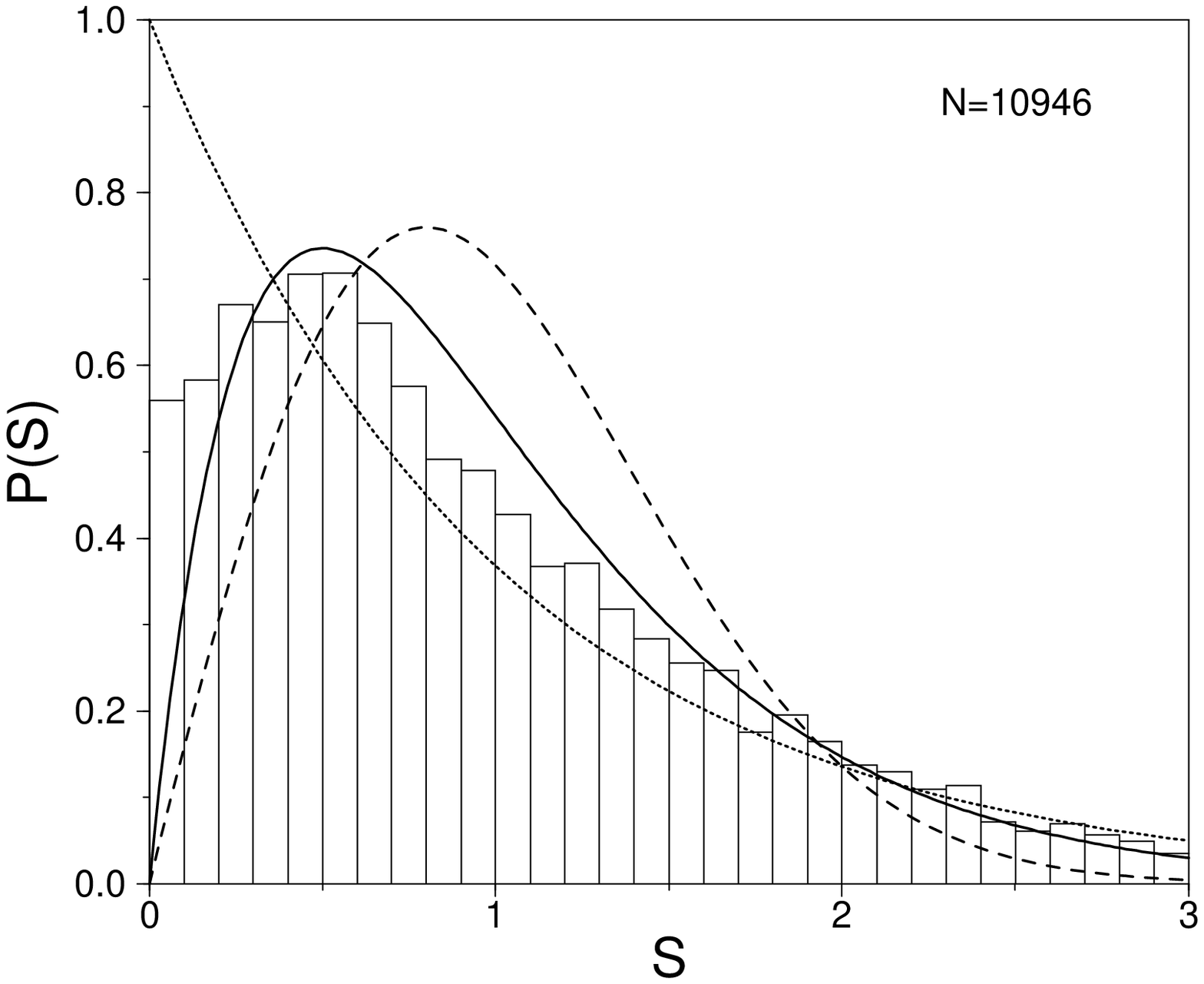,width=8.0cm}}
{\footnotesize{{\bf FIG. 3.} {\bf (a), (b)} 
The calculated nearest level-spacing distribution $P(S)$ 
by unfolding the data for $\rho(log(E))$ for two Fibonacci matrices
of sizes $N=N_{17}=4181, N_{19}=10946$.
The dashed line is the Wigner surmise $P(S)=(\pi/2)Sexp(-(\pi/4)S^{2})$
(extended states), the dotted line is the Poisson law 
$P(S)=exp(-S)$ (localized states)
and the solid line is the semi-Poisson distribution 
$P(S)=4Sexp(-2S)$ (critical states). The asymptotic distributions for 
extended and localized states are less relevant for our 
system since the data seem to approach the semi-Poisson 
law for the sizes used.
}}
\vspace{.3in}
%
%%%

The obtained spacing distribution shown in fig. 3 is neither Wigner
(extended) nor Poisson (localized). It is better described by a
semi-Poisson intermediate level-statistics known to characterize
pseudointegrable billiars\cite{6},
disordered systems at the critical point after performing certain
average over boundary conditions \cite{7}, 
quasiperiodic critical systems \cite{8,9},  
interacting particles \cite{4}, etc.  
Although the agreement with the semi-Poisson 
curve $P(S)=4Sexp(-2S)$ is  not perfect the obtained curve
shows some scale-invariant characteristics  and we can
attribute the small disagreement with the semi-Poisson curve
as due to the rather small number of levels considered.
For example, the maximum size of $10946$
considered in our computations, which gives the same number of levels,
is clearly insufficient to answer definitely this issue,
since  over $100000$ levels are usually required for this purpose
\cite{7,8}, clearly impossible at present. 
However, we expect the agreement to improve as $N$ increases.
The discussion up to now also answers the second question.

%%
%%-------------------fig 4a,b
%%
\par
\vspace{.3in}
\centerline{\psfig{figure=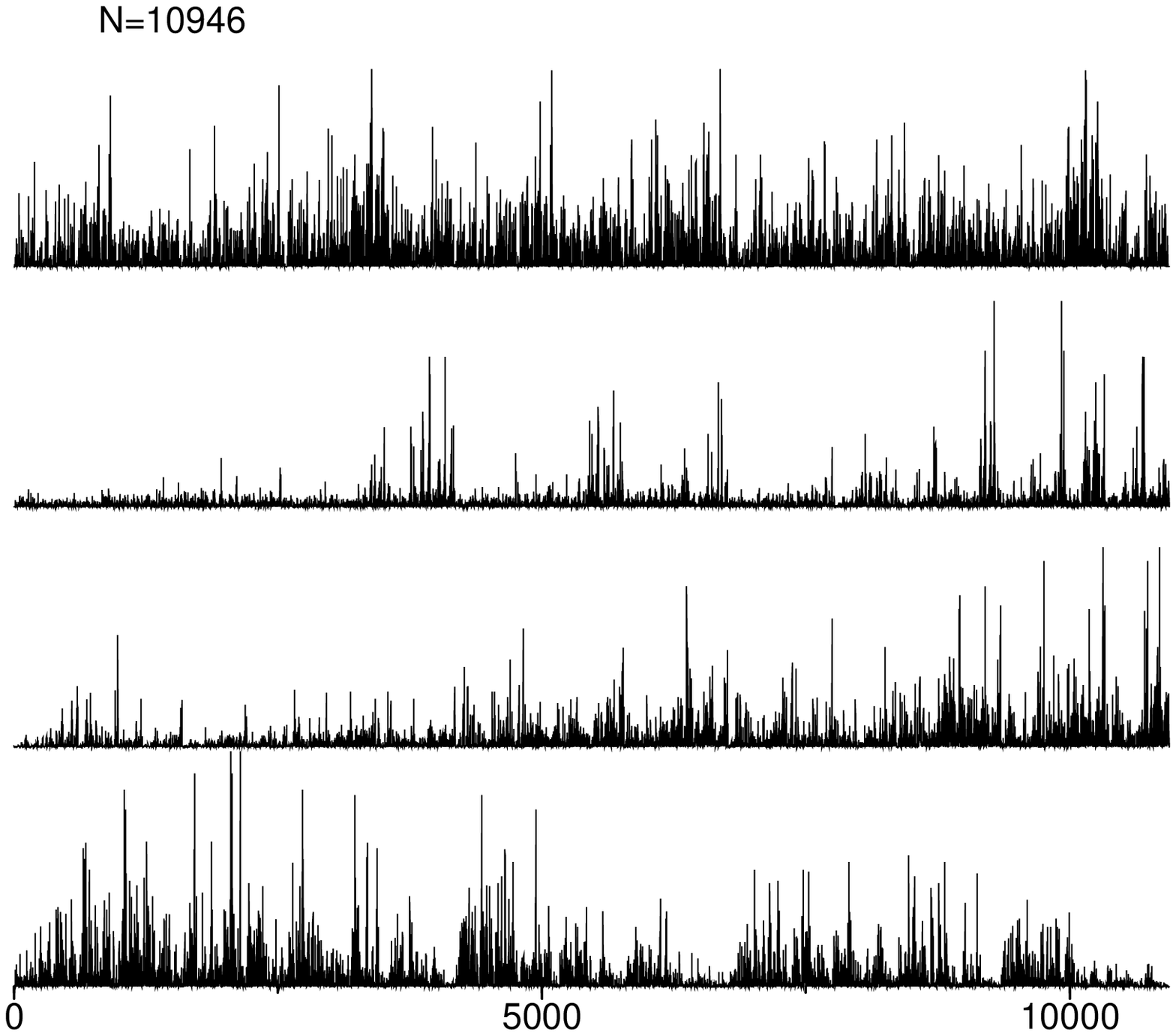,width=8.0cm}}
\centerline{\psfig{figure=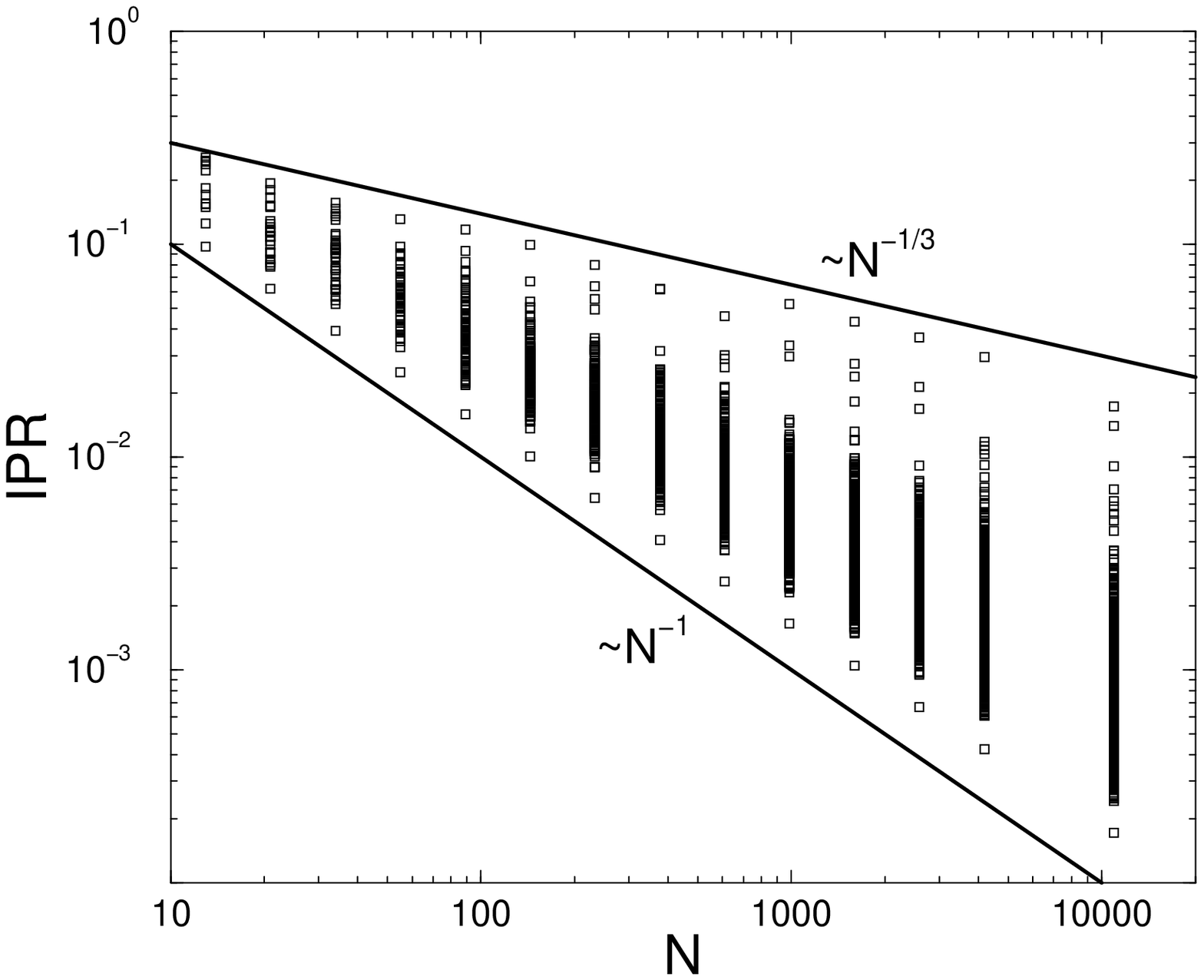,width=8.0cm}}
{\footnotesize{{\bf FIG. 4.} {\bf (a)} 
The computed typical eigenvector amplitude distributions for 
the Fibonacci matrix of size $N=N_{19}=10946$.}
{\bf (b)} The corresponding scaling of the inverse participation 
ratio vs. size $N$ for all eigenstate amplitudes, which
defines their $D_2$ fractal exponents to lie between $1/3$ and $1$.}
\vspace{.3in}
%
%%%

Finally, in order to answer the third question 
we have considered the eigenvector behavior of the Fibonacci
matrix. Results for some typical eigenvector amplitude distributions 
together with the scaling of the inverse participation ratio 
are shown in fig. 4.
We find that the vast majority of the eigenvectors are `critical' 
sharing delocalized fractal characteristics.  The scaling
of the inverse participation ratio allows to compute $D_{2}$ 
exponent for the eigenvector amplitudes. We obtain
$1/3\leq D_{2}\leq 1$ while only very few eigenstates of this matrix
are truly extended with $D_{2}=1$ usually lying at the 
spectral edges while states close to the band center are 
more `localized' ($D_{2}\approx 1/3$).

In summary, we have obtained numerically the full eigensolutions 
of a fractal Fibonacci matrix. The results show critical level-statistics 
with almost semi-Poisson nearest-level distribution 
and multifractal eigenvectors.
The Fibonacci matrix studied shares similar characteristics to 
one-electron disordered systems at the mobility edge, quasicrystaline and 
also interacting electron systems. Moreover, it can be interesting
for quantum systems which can show intermediate level-statistics 
of what is believed to be `critical chaos', in the border between 
quantum chaos and integrability.
In might also offer some hints for the expected behavior from the 
diagonalization of matrices in complicated many-body Hamiltonians.

%%

%\end{multicols}
\end{document}